\documentclass[twocolumn,aps,pra,amsfonts]{revtex4-2}
\usepackage{graphicx} % Required for inserting images

\newcommand{\ket}[1]{\left|#1\right\rangle}

\newcommand{\braket}[2]{\left\langle{#1}|{#2}\right\rangle}

\begin{document}
\title{How to be a Copenhagenistic-QBistic Everettist?}
\author{Marcin Wie\'sniak}
\affiliation{Institute for Theoretical Physics and Astrophysics, Faculty of Mathematics, Physics, and Informatics,\\ University of Gda\'nsk,  80-308 Gda\'nsk}
\date{\today}

\begin{abstract}
    This essay reviews a modern understanding of a quantum measurement. Rather than reducing the picture to the observer's experience with quantum system, we try to put it in the context of a broader physical picture. We also attempt to distinguish some basic components and stages of a generic measurement. In this way, we want to understand which aspects of quantum measurements are important to mainstream interpretations of quantum mechanics.
\end{abstract}
\maketitle
\section{Introduction}
Since its birth, quantum mechanics has always caused problems with its interpretation and compatibility with other sciences. This problems originate in a different status of an agent (an observer) and an measurement. In other sciences, the agent collects information about the subject without inferring its properties. It is only at a quantum level that the observer significantly perturbs the system. Moreover, only in quantum mechanics answers to some questions are restricted to a countable set and appear in a random manner

The questions that immediately follow from this fact are, for example, if the agent does indeed know everything possible about the system, which would cover the randomness issue, if interactions are relativisticly causal, or if the agent is described by the same physical principles as the system. If so, then how is she distinguished to be able to conduct a measurement, but if it is not the case, where is the demarcation line. Collections of answers to questions such as these were gathered in various interpretations of quantum mechanics (e.g. Refs. \cite{bohm1952suggested,everett1957relative,watanabe1955symmetry,ghirardi1986unified,cramer1986transactional,zurek2009quantum,fuchs2010qbism}).
Each of these interpretations addresses certain philosophical questions, but ignores some other. In this essay we do not attempt understand quantum physics from the foundational point of view, but rather by basing on our understanding of other research fields, including electromagnetism and elements of physiology. Specifically, we want to look in details at the act of a measurement. While such as analysis still leaves some fundamental issues unanswered, this approach allows us to understand a bit better which features are highlighted by which interpretations and where are the transitions between them.
\section{Cyanobacterium evolves to quantum physicist}
Cyanobactreria, the first photosynthetizers on Earth, might appeared in waters about 3.2 billion year ago \cite{des2000did}. This event may mark the first association of physical, rather than chemical, conditions to the survival strategy. The first predators may have appeared as early as 2.7 billion years ago  and first complex animals evolved c.a. 2.1 billion years before today \cite{el20142}. The relation between intensity of ambient light and nutrition availability, to both prey and predators, must have affected these early lifeforms. As evolution progressed, more complicated survival strategies basing on external conditions were necessary to develop. For example, a biological cycle had to be adjusted to days rich in solar energy and nights, when resources needed to be spared. Likewise, as plants conquered dry lands, they had to adjust to harsh winters, for example by rejecting their photosynthesis organs (leaves) and rely on accumulated supplies, or to frequency of rains.

Eventually, finding patterns in external stimuli has become a crucial part of survival strategy and as such it was linked to the mesocorticolimbic circuit (the reward system) giving rise to scientific curiosity. Significance of these patterns to our existence, a general agreement in their interpretation (from dropping leaves in autumn to using Schr\"odinger's equation for predictions) and the ability to learn about them from other individuals highlight their external nature and objectivity. Otherwise, why would the agent's mind generate cryptic information relevant for its continuity and disguise its meaning in imaginary characters? It is rather safe to assume that there is a physical reality governing our experience and our existence is dependent on its laws.

A physical picture of the Universe was, up to the beginning of the 20th century, fairly consistent and comprehensive, basing on Newtonian mechanics,
electrodynamics and statistical mechanics. However, a closer inspection of both theoretical predictions and more precise experiments caused cracks in this picture, vividly discussed to this day. On one hand, properties of electromagnetic waves gave a rise to special relativity \cite{einstein1905elektrodynamik}, which nivelated a distinction between space and time dimensions. Introduction of special relativity \cite{einstein1911einfluss} to a refutation of the hypothesis of etherium, a misterious medium for light. Later, general relativity was introduced, defying the absoluteness of space-time, as it becomes curved in presence of, now equaled, energy-matter. An example of unobvious feature of general relativity are black holes \cite{schwarzschild1916gravitationsfeld}, inaccessible regions of spacetime with infinite curvature, now well evidenced \cite{collaboration2019first}. 

On the other hand, the understanding of light and the atomic structure has led to a conclusion that certain physical quantities must come in portions, quanta. This complicates, rather than simplifies, the description of a system, as it typically involves more than one classical state. This coexistence, known as the superposition, does not have a proper analogue in logic, but establishes a new relation parameterized by complex numbers. It even extends to complex systems consisting two or more parts, allowing for entangled states. States from this class exihibit correlations stronger than obtainable with local operations and classical communication \cite{schrodinger1935gegenwartige}. In general, the probability of observing a system described by final state $\ket{f}$, given that its initial state is $\ket{i}$ is governed by the Born rule, 
\begin{equation}
    P(\ket{i}\rightarrow\ket{f})=|\braket{i}{f}|^2
\end{equation}. 
Pure states $\ket{i}$ and $\ket{f}$ can be either normalized vectors with some $d$ complex components or complex functions integrable to 1 in squares of moduli, and $\braket{\cdot}{\cdot}$ represents thus scalar product of them. A more general description of states, which allows for an incomplete knowledge about the system is based on mixed states, normalized, positive semi-definite matrices or operators.

For the very first time, the classical picture of reality accumulated over the history of science has been shaken. From above -- in the limit of extremely high speeds, masses and distances, from below -- by behavior of the smallest building blocks, in a realm where an act of observation itself influences the experiment. Soon after it turned out that these two extreme regimes can even be incompatible with each other, with examples of difficulties in formulating fully relativistic quantum mechanics or quantized gravity theory, the problem of expansion of the Universe, or the breakdown of quantum formalism in the limit of infinitely many subsystems.

It should come as no surprise that these developments were not immediately accepted by the general public. This was the case especially with quantum theory, even among its founding fathers. Schr\"odinger proposed a gedanken experiment with a cat in a sealed box to be driven to a superposition of being alive and dead to ridicule predictions of the new theory \cite{schrodinger1935gegenwartige}. Einstein struggled with accepting randomness of quantum mechanics and with together with Podolsky and Rosen \cite{einstein1935can} postulated supplementing quantum theory with additional parameters intrinsically hidden from the observers. This work later stimulated the discussion on hidden variable (HV) models. The conclusion reached today seems to be that while theories with general HV models are plausible \cite{bohm1952suggested}, specific variants, such as local (LHV) \cite{bell1964einstein}, a specific kind of nonlocal (NLHV) \cite{leggett2003nonlocal}, or non-contextual (NCHV) hidden variable models \cite{kochen1990problem} are falsified by theoretic works supported by accompanying experiments (e.g. Refs. \cite{shalm2015strong,hensen2015loophole,giustina2015significant}). Freedman and Clauser conducted their pioneering experiments on Bell's inequalities \cite{freedman1972experimental} with an expectation that they will actually display obedience of the limits imposed by LHV theories.
\section{Quantum physicist measures system}
In quantum mechanics an agent learns about the surrounding through an act of measurement. The agent interacts with the system in a controllable way to relate to one of its properties and, as she believes, receives a one-time definitive answer. Almost always such an experiment must be repeated a number of times to establish some statistics. An act of measurement is associated with either a destruction of the system or an update on future predictions concerning it. Once quantum mechanical formalism of state vectors and observables is accepted, certain pairs of measurements are be incompatible by the Heisenberg uncertainty principle. They cannot be faithfully conducted simultaneously or in sequence as one of them disturbs the outcome of the other.

These features of quantum mechanics called for an interpretation, which would bring it back closer to the legacy of empirical sciences. Such an interpretation should address questions such as ``what is a measurement?'', ``who or what is the agent and what is her role in the act of measurement?'', ``why does the agent report only one outcome of the measurement?'', or ``what can we deduce about the underlying structure?''. As quantum mechanic was the first inevitably random theory, a particularly interesting matter for these interpretations was ``is there a reason that this particular result and not any other was reported in a measurement?'', effectively reviving the HV discussion. The most highlighted interpretations are as follows.

{\em The Copenhagen interpretation}, in which an observer conducts a measurement of her choice. The system responds by a reduction of its initial state and yielding an output through the measuring device so that the new state describes the measured property accordingly to the observed outcome. There is a clear distinction between classical macro-reality of the agent and quantum micro-reality of the system.

{\em QBism} \cite{fuchs2010qbism} or quantum Bayenisianism is a radical minimalistic philosophical point of view, which only acknowledges the agent and her experiences. As these experiences accumulate, the agent updates her predictions about future events, eventually reaching the Born rule.

{\em Everett's} interpretation \cite{everett1957relative} states that at the (rather loosely defined) moment of measurement, the Universe branches to many its copies, each described by a wave function corresponding to a different outcome. This does not necessarily mean that the Universe becomes physically multiplied, but that there are multiple timelines open.

In the catalogue of interpretations of quantum mechanics we find various other proposals, many of which fail to pass the William of Ockham razor criterion, i.e., introduce new entities or features only to give an explanation. Some of them may also lead to paradoxes and contradictions, drawing us away from the desired comprehension. For this reason one may focus on the above three. For example, TSFV \cite{watanabe1955symmetry} and the transactional interpretation \cite{cramer1986transactional} of quantum mechanics introduce advanced and retarder wave-functions to the description, without specifying what ends an experiment. Moreover, at least TSVF does not obey the Born rule.

To provide some understanding of a measurement in quantum mechanics, it could be beneficial to decompose it to more elementary stages. A more detailed description may differ depending on the type of the system, the measured quantity and the measuring apparatus, but one possible framework is described in what follows.

A quantum measurement is an intentional arrangement of evolution that leads to a correlation between the measured property of the system and an internal state (knowledge) of the agent. Note that decoherence is a broader notion than a measurement. A system could couple to orthogonal states with identical photon statistics, which would not be directly accessible to the agent. For the same reason it is not enough to require spectrum broadcasting from the evolution, but that many consistent signatures of each pointer states are broadcast \cite{zurek2009quantum}.

A measurement will involve a few entities. First, an important part is the agent herself. We also need to distinguish some quantum system we want to study, and some instrument to do it with. This instrument will have some richer inner structure. We also assume some medium, which is a system coupled both to the measuring apparatus and the agent. Finally, we will use the concept of an unspecified environment.

The first stage is {\em transformation}. Because we are only able to register events well localized localized in spacetime, most often the agent needs to couple the quantity she wants to study to the position or times of arrival to the measurement device. In technical terms, at this stage we map the desired pointer basis to a subbasis of localized states. In optics this could be realized by a system of mirrors and phase plates, for spin measurements it is rather a spatial arrangement of magnets, etc.. Importantly, this stage expresses the will of the agent about what is measured. Technically, these transformations are nearly perfect realizations of unitaries of the system, and therefore they should be considered as reversible. However, as measurements are not necessarily compatible with one another, this is the stage where the agent may execute her free will, by arragning one and not another setup.

We can then distinguish {\em conversion}, during which the system interacts with a detector. The role of detector, as understood here, is to provide interface between the system and some signal that can be accessible to agent. A textbook example is a moment when a photon hits the first cathode in a photon multiplier. A more modern case is that of transition edge detectors, where the system excites a phonon in a SQuID, which results in a change of a magnetic flux. Typically, the system-detector interaction is very weak, resulting in minuscule probability of reaction. To increase the probability, but also for manufacturing convenience, an experiment would rather involve an array of such elementary detectors.

Immediatelty after the conversion starts {\em amplification and broadcasting}. It does not need to be in terms of energy of the signal, though most often energies associated with the system are to low for the agent More importantly it is amplification of coherence, transforming a superposition into a multipartite entangled state. If a superposition of states of the system was an applicable description up to this point, now it refers to states describing also the medium, by which the agent will be communicated, e.g., light, and other factors. For example, optoelectronic sensors include actual amplifiers (often transistor-based) for photoelectrons. A grain on a photosensitve layer absorbs or scatters photons falling onto it. It was pointed out, e.g., by \.Zukowski and Markiewicz \cite{zukowski2024against} that this amplification does not need to couple orthogonal states of the system/detector with orthogonal states of the medium. Consider the following interaction:
\begin{eqnarray}
\label{COUPLING}
&&\ket{0}_S\ket{0}_M\rightarrow\ket{0}_S\ket{0}_M,\nonumber\\
&&\ket{1}_S\ket{0}_M\rightarrow\ket{0}_S(\cos\alpha\ket{0}_M+\sin\alpha\ket{1}_M).
\end{eqnarray}
Choosing, say $\alpha=\pi/3$ and applying Eq. (\ref{COUPLING}) to only ten different systems provides the squared modulo of the overlap between the two states of the medium, i.e., the probability to confuse the two, is $9.55\times 10^-7$. 

It shall be pointed out that in practice the medium almost always involves the electromagnetic field. A peculiar example can be that of a Geiger-M\"uller counter, where a detection was primarily announced to the agent by an audible click, but it originates in electric discharge in a gas. The electromagnetic radiation is naturally omnidirectional, and it propagates with the speed of light, making it particularly difficult to control in natural arrangements. Involving the electromagnetic field itself greatly contributes to the effective impossibility of a precise theory of measurement.

The final stage of the measurement process is {\em perception}. It starts at the moment that information in the medium reaches the agent to stimulate her senses. In this way she gets coupled to the system. On the other hand, perception last as long as the agent contemplates her correlation to the system in this particular act of measurement.

In general, it is at this stage, where the outcome statistics are broadcast to the environment. Branches of the state corresponding to different observed results are separated by very large Hamming distances, i.e, distinguishably different states of many subsystems. The agent receives only a small fragment of the amplified state, but still significant enough that she can distinguish between the resulting events.

\section{System does and does not get a description}
It is a philosophical debate whether the evolution of the Universe  can be described by some global wave-function $\Psi$ the Schr\"odinger equation,
\begin{equation}
    i\hbar\frac{\partial}{\partial t}\Psi=H\Psi,
\end{equation}
where Hamiltonian $H$ could be, for example, of Dirac's type, but any physical equation is only as good as its predictions. In practice, tracking the evolution of dozens, not to mention thousands of particles with their positions, charges, internal states and potentials requires restricting oneself to statistical method.  

First, let us ask which stages may contribute to creating correlations between the system and the rest of the Universe. By definition, transformation is a unitary operation on the system, which means that it can be undone by using the same device. Naively, we would believe that conversion by the detector can also be seen as coupling of two relatively simple systems, but typically ``elementary detectors'' appear in a large numbers -- to increase the collection rate and due to convenience of manufacturing.

 In contrast to transformation, amplification is specifically defined as correlating the system (through a detector) to as many particles in the medium as possible. Note that the medium is is not designed to focus directly on the agent but it rather spreads across the whole space. The agent only receives a small part of the medium, the vast majority of it propagates in void or to other agents, leading to intersubjectivity of observations. In perception, the signal reaches the agent, but this process now involves the complete mechanism of transferring information in the nervous system. Using concentric cables still would not be a good approximation of a lossless electronic communication, but biological systems are based on chemical computing. An electrical signal is transmitted through neurons by modulating calcium and potassium ions concentration with a cell membrane. The necessary inclusion of these objects in the description puts us even farther apart from fully understanding the measurement process. Note that, like any other organ, the brain constantly emits energy stored in chemical resources as a thermal radiation, which can still contain some information about the perception of the outcome.

A minimal light intensity that can cause a reaction in a human brain is estimated to be five to nine photons within a time window of 100ms. Producing entangled states of that many photons is well within the reach of modern quantum optical experiments. However, once the state would reach the agent's eyes, all entanglement would be immediately lost due to extremely convoluted mechanics of biological systems. Information becomes classical immediately upon reaching the agent, at the very latest. This leaves no space for any practical realization of the Schr\"odinger's cat \cite{schrodinger1935gegenwartige} or the Wigner's friend \cite{wigner1995remarks} gedanken experiments. Just like in the case of the Universe, the unitary description of Wigner's friend(s) may be still formally correct, but it is not feasible and therefore not relevant, anymore. 

Modern everyday computers allow to handle arbitrary mixed states of about 14 qubits (two-dimensional quantum systems). Supercomputers could deal with probably not much more than 30 qubits. That is to say, these machines can hold such states in memory and do some simple operations on them, e.g., calculate a mean value of some operators. The main obstacle in attempting a more general modelling is that multiplying two $d$-dimensional matrices requires $O(d^3)$ floating point operations.

Computational and operational complexity of the problem will always be an argument that reversibility of quantum measurements is not attainable {\em in practice} rather than {\em in principle}.  As it seems, however, the analysis cost escalates so rapidly beyond any practical feasibility that it is safe to say that a reversal of a measurement is impossible {\em in principle}.

A similar approach to the measurement problem was initialized already in Ref. \cite{zeh1970interpretation}. However, Zeh, primarily focused on chaotic evolution within the measurement device. He believed that the main mechanism for decoherence was phase randomization. On the other hand, Zurek \cite{zurek2009quantum} focuses on the state of the environment and is concerned about methods for (exact) coherence amplification. Here, we try to include all stages of the measurement and associate decoherence for a vast plethora of possible states involved in the measurement, so much so, that knowing phases between them is not that relevant anymore.

A Copenhagenist or a QBist would welcome the acknowledge the distinguished of the agent. It is currently, and perhaps always be, unfeasible combine quantum mechanics with what we consider a free will or consciousness, so we rather postulate these attributes. Moreover, it is in spirit of these two interpretation that a measurement would involve a will of the agent.

Nodding to the Everett's interpretation we observe that our model of measurement does not include any model for physical collapse. The agent experiences different measurement outcomes in different timelines, which may as well continue to infinity. In each timeline there is a consistent historical record leading to the present moment. However, as there is no further connection between them due to an extremely complex evolution, we experience only one timeline and others are only hypothetical.

What about the state: is it a property belonging to a quantum system, or a mere mathematical tool? The Everett and, to some degree, the Copenhagen interpretation stress the physical relevance to the system. Consider a nondestructable quantum entity, for example, a spin of a particle. At some point of time we describe it with a pure state, but we could only draw such a statement from a measurement. A Copenhagenist then refers to a collapse, but as we discussed, there is no physically motivated model for it. In Everett's and QBistic interpretations, rejecting a hypothesis that our spin never interacted with any other part of the Universe, we have to assume that the state is a Bayesian concept referring to our knowledge, updated only at some recent time with our experience.  Therefore, it does not describe any correlations that spin could have acquired earlier.
Also, we can ascribe states to yet nonexisting systems, such as photons from a pulsed source. As QBists stress, we do it basing on our previous experience, mastering the betting strategy.

Finally, one could ask about the underlying physical reality. QBism refuses to discuss any physical realiżation focusing specifically on experience of the agent. This approach seems to be convergent with the point of view of, e.g., Zeilinger, who occasionally expresses his opinion that ``photons are just clicks in a detectors'' \cite{arndt2005quantum}. Indeed, it is impossible to infer the nature of the underlying reality basing on the experience. However, taking this observation as an imperative is too radical. It defies the essence of science, which is to find relations between our observation in order to make some predictions about future events. Building these relations require a language to describe them. The language of mathematics provides necessary technicalities, but lacks the essence. A way of describing phenomena as a series of collisions of particles proved itself to be exceptionally versatile and useful, even if not complete. Thus particles exist at least at the same level as numbers -- as concepts. Likewise, we should accept the evolution of these particles as given. Contrary to QBism's denial, physical laws should apply regardless of the role of the agent. On the other hand, an inspection of a measurement process does not lead to the Born rule, yet it is, in a QBist's words, the best betting strategy. We cannot know what the physicality is, only how it works.

\section{Conclusions}
Quantum mechanics has always posed challenges in our understanding of the surrounding world. One of the most notorious question is why that of a measurement: how does it happen and what happens to its result. In order to comprehend this problem, a number of interpretations were proposed, from as minimalistic a QBism \cite{fuchs2010qbism} to those introducing new physical entities and mechanisms, as in case the GRW objective collapse model \cite{ghirardi1986unified} or the two-state vector formalism \cite{watanabe1955symmetry}. In this essay we accepted the common understanding of physical evolution and attempted to decompose a use of a measuring device into bit more fundamental stages. Such an analysis gives no rise to any physical mechanism of a wave-function collapse, but rather highlights a rapid loss of coherence between classically distinguishable states. The loss occurs already at the level of the measuring apparatus and prevents the observer to experience multiple outcomes. Even if the agent is fed directly with a quantumly correlated signal, it is immediately reduced to classical information. 

It is therefore not the real problem of measurement to answer why a particular outcome was observed in each run, as we argue that every possible outcome occurs. As we have stressed above, we have no means to judge about the underlying reality, but we ought to give a possibly consistent story explaining as many phenomena as possible. The real challenge, probably never to be fully answered, is about the agent, how she is able to plan measurements and predict their results.

Different interpretations of quantum mechanics do not originate in disagreements, but in emphasizing various aspects of the same phenomena. Therefore we should not spare efforts in their unification.
\section{Acknowledgements}
This work has been suported by NCN (Poland) Grant 2017/26/E/ST2/01008.
\bibliography{cite.bib}
\end{document}